# Imaging electron flow from collimating contacts in graphene

**S Bhandari[†], G H Lee[&*], K Watanabe[ᵍ], T Taniguchi[ᵍ], P Kim[†&], R M Westervelt[†&]**

[†]School of Engineering and Applied Sciences, Harvard University, MA 02138, U.S.A
[&]Department of Physics, Harvard University, MA 02138, U.S.A
[ᵍ]National Institute for Materials Science, 1-1 Namiki, Tsukuba, 305-0044, Japan
[*]Department of Physics, Pohang University of Science and Technolgy, Pohang 790-784, Republic of Korea

Email: sbhandar@fas.harvard.edu

**Abstract**. The ballistic motion of electrons in graphene encapsulated in hexagonal boron nitride (hBN) promises exciting opportunities for electron-optics devices. A narrow electron beam is desired, with both the mean free path and coherence length exceeding the device size. Shaping the electron beam with patterned gates does not work in monolayer graphene, because the conductivity cannot be driven to zero due to the absence of bandgap. One can form a collimating contact in graphene by adding zigzag contacts on either side of the electron emitter that absorb stray electrons to form a collimated electron beam [23]. Here we provide images of electron flow from a collimating contact that directly show the width and shape of the electron beam, obtained using a Scanning Gate Microscope (SGM) cooled to 4.2 K. The device is a hBN-encapsulated graphene hall bar with narrow side contacts on either side of the channel that have an electron emitter at the end and absorbing zig-zag contacts at both side. To form an image of electron flow, the SGM tip is raster scanned at a constant height above the sample surface while the transmission to a receiving contact on opposite sides of the channel is measured. The SGM tip creates a capacitive image charge in the graphene sheet that deflects electrons away from their original paths, changing the transmission $T$. By displaying the change $\Delta T$ vs. tip position, an image of ballistic flow is obtained. The angular width of the electron beam leaving the collimating contact is found by applying a perpendicular magnetic field B that bends electron paths into cyclotron orbits. SGM images reveal that electron flow from a collimating contact disappears quickly at B = 0.05T, because the electron paths are bent away from the receiving contact, while the flow from a non-collimating contact persists up to $B$ = 0.19 T. Ray tracing simulations agree well with the experimental images over a range of $B$ and electron density n. By fitting the half-width at half-maximum (HWHM) of the magnitude of electron flow in the experimental SGM images, we find a narrow half angular width $\Delta\theta = 9.2°$ for the electron flow from the collimating contact, compared with a wide flow $\Delta\theta = 54°$ from the non-collimating contact.

## 1. Introduction:

Ballistic graphene devices open pathway for new electronic and photonic applications [1-6]. In addition, electrons in graphene move at constant speed $\sim 10^6$ m/s and they pass through potential barriers *via* Klein tunneling, due to the gapless bipolar nature of monolayer graphene [7-11]. Such differences in graphene from conventional 2DEGs offer challenges to understanding the device physics, and the design of new devices that make use of graphene's virtues. For ballistic electronics, a narrow, collimated electron beam, with all the electrons pointed in nearly same direction would be ideal. The motion of electron beam through a ballistic device could be controlled, by bending the paths with magnetic fields, attracting or repelling them with electric fields and electron mirrors [12-17]. The formation of a collimated beam in a two-dimensional electron gas (2DEG) in GaAs has been previously observed in flow between two separated quantum point contacts (QPCs) [18], and focusing has been achieved by a magnetic field [19] and an electrostatic lens [20]. The angular distribution of the lowest quantum mode of a QPC provides a degree of collimation [21, 22] but a more tightly directed electron beam is desirable. Previously, collimation of electron flow using two constrictions in graphene was demonstrated via transport measurements [23].

In this paper, we demonstrate that a collimating contact can produce a narrow electron beam in graphene by imaging the electron flow using a cooled Scanning Gate Microscope (SGM). We have adapted a technique previously used to image electron flow through a GaAs 2DEG [16,21,22] to graphene [17,24] and used this technique to image electron flow from a collimating contact in graphene. The collimating contact is formed by a rectangular end contact with zigzag side contacts on either side that form a collimated beam of electrons. Images of electron flow confirm that the collimating contact substantially narrows the electron beam. A quantitative measure of angular width of the electron beam is obtained by applying a perpendicular magnetic field $B$ to bend the electron trajectories into cyclotron orbits, so they miss the collecting contact, reducing the intensity of electron beam image. A fit to ray tracing simulations gives a (HWHM) angular width $\Delta\theta = 9.2°$ for the collimating contact and much wider width $\Delta\theta = 54°$ when collimation is turned off.

## 2. Methods

*2.1 Collimating Contact Device:*

The geometry of the collimating contact is shown in Fig. 1a, an SEM image of the Hall-bar graphene sample; the white square indicates the imaged region. The Hall bar (blue region) is patterned from a hBN/graphene/hBN sandwich. It has dimensions 1.6x5.0 $\mu m^2$, with two collimating contacts (yellow) along each side, separated by 1.6 $\mu$m, and large source and drain contacts (width 1.6 $\mu$m) at either end. The four collimating contacts have an end contact that emits electrons and two zigzag side contacts that collimate the electron beam by absorbing electrons. The device sits on a heavily doped Si substrate which acts as a back-gate, covered by a 285 nm thick insulating layer of silicon oxide ($SiO_2$). The top and bottom BN flakes along with the graphene are mechanically cleaved. Using a dry transfer technique, the flakes are stacked onto the $SiO_2$ substrate. To achieve highly transparent metallic contacts to the graphene, we expose the freshly etched graphene edge with reactive ion etching and evaporate chromium and gold layers immediately afterwards [25].

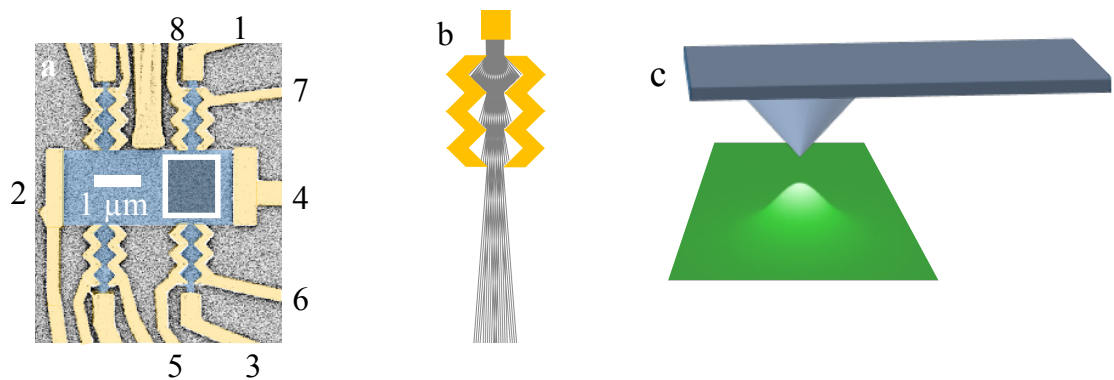

Figure 1: (a) Scanning electron micrograph of hBN-graphene-hBN device in a hall bar geometry with four collimating contacts with absorptive ziq-zag sections that form a narrow electron beam. The white square shows the area imaged by the cooled scanning gate microscope. The blue region indicates graphene and yellow indicates metal contacts(b) Ray-tracing simulations of electrons passing through the collimating contact (yellow) through absorptive zig-zag side contacts (yellow). (c) Simulated image charge created by the charged tip which creates a local dip in electron density.

Figure 1b shows ray-tracing simulation of electron trajectories passing through a collimating contact, which consists of an end contact (yellow) that emits electrons into the graphene and zigzag side contacts on either side (yellow) that form a series of constrictions. Electrons emitted from top narrow contact enter at all angles. Collimation is turned on by grounding the zigzag side contacts - stray electrons entering at wide angles hit the zigzags and are absorbed - only electrons that pass through the gap get through, producing a narrow electron beam. The collimating contact can be turned off by simply connecting the top contact to the two zigzag side contacts. In this case, the combined contact behaves as a single source of electrons with a wider width and no collimation.

*2.2 Cooled Scanning Gate Microscope*

We have developed a technique that uses a cooled Scanning Gate Microscope (SGM) to image the flow of electrons through a two-dimensional electron gas that we used for GaAs/AlGaAs heterostructures [16,21,22] and graphene samples [17,24,26]. The charged tip creates an image charge in the 2DEG below (Fig. 1c) that deflects electrons away from their original paths, changing the transmission $T$ between two contacts of a ballistic device. An image of the electron flow is obtained by displaying the change in transmission $\Delta T$ as the tip is raster scanned across the device.

In this paper, we used this approach to image electron flow from a collimating emitter contact at the top of the graphene sample in the area indicated by a white square in Fig. 1a to a non-collimating collector contact at the bottom of the sample. With the tip absent, electrons pass ballistically through the channel between the emitter and the collector contacts. The shape of the electron beam is imaged by displaying the change in transmission $\Delta T$ *vs*. tip location as the SGM tip scatters electrons away from their original paths, as the tip is raster scanned across the sample. We measure voltage difference $\Delta V$ between (1, 7, 8) and 4. One can determine $\Delta T$ from the change in voltage $\Delta V$ at the ungrounded collecting contact for an current $I$ into the emitting contact. As electrons accumulate, raising the electron density, the chemical potential increases, creating an opposing current that maintains zero net current flow. By measuring the voltage change $\Delta V$ or the transresistance change $\Delta R_m = \Delta V/I$ at the receiving contact, the transmission change $\Delta T$ induced by the tip can be obtained [17,26]. For a collimated electron beam, the current is emitted from contact 1 to contact 2 in Fig. 1a, while contacts 7 and 8 are grounded to collect sideways moving electrons. To turn collimation off, contacts 1, 7 and 8 are connected together as a single current emitter while contact 2 is grounded. The collecting contact always has collimation turned off with contacts 3, 5 and 6 connected together.

The width of the emitted electron beam is substantially reduced when collimation is turned on for the top contact. To obtain a quantitative measure of the angular width of the emitted electron beam, we apply a perpendicular magnetic field $B$ that bends electron paths into cyclotron orbits. The curvature causes electrons to miss the collecting contact and reduces the intensity of the imaged flow.

*2.2 Electron path simulation:*

Bending electron trajectories with a perpendicular magnetic field $B$ is used to measure the angular width $\Delta\theta$ of the emitted electron beam in the experiments below. We use a ray-tracing model of electron motion and tip perturbation to simulate the electron flow through graphene in this case [17,26]. Our model computes the transmission of electrons between two contacts in graphene for each tip position.

The electrons trajectories from the emitter contact are traced by considering two forces: 1) the force from the tip induced charge density profile, and 2) the Lorentz force from *B*.

The work function difference between the Si SGM tip and graphene creates a change in electron density $\Delta n_{tip}$ in the graphene below the tip. For a tip with charge *q* at a height *h* above the graphene sheet, the change in density $\Delta n_{tip}$ at a radius *a* away from the tip position is :

$$\Delta n_{tip}(a) = \frac{qh}{2\pi e(a^2 + h^2)^{3/2}} \qquad (1)$$

where *e* is the electron charge. We choose a peak density change $\Delta n_{tip}(0) = -5 \times 10^{11}$ cm$^{-2}$ at *a* = 0 to match the experimental data. The change electron density $\Delta n_{tip}(a)$ results in a change in the Fermi energy $E_F(a) = E_F(n + \Delta n_{tip})$. The total chemical potential, given by $E_F(a) + U(a)$ where $U(a)$ is the electrostatic potential due to capacitive coupling to the tip, must be constant in space. Taking a spatial derivative yields the force $F(a) = -\vec{\nabla}U(a) = \vec{\nabla}E_F(a)$ on electrons in graphene passing nearby the tip position. In graphene, the Fermi energy is $E_F = \hbar v_F(\pi n)^{1/2}$ and the electron dynamical mass is $m^* = \hbar(\pi n)^{1/2}/v_F$ [17,26,27] so the acceleration of an electron due to the tip at position $\vec{r}$ is:

$$\frac{d^2\vec{r}}{dt^2} = \frac{1}{2}\left(\frac{v_F^2}{n}\right)\nabla n(\vec{r}) \qquad (2)$$

The tip-induced charge density profile creates a force that pushes an electron away from region with low electron density beneath the tip. The Lorentz force *F* that acts on an electron with velocity *v* under a magnetic field B is:

$$\vec{F} = e\,\vec{v} \times \vec{B} \qquad (3)$$

In our simulations, we pass *N* = 10,000 electrons at the Fermi energy into the graphene from the emitting contact. The number of electrons passed from the contact follow a cosine distribution where maximum number of electrons pass perpendicular to the contact. The distribution is cosine within the angular width $\pm\Delta\theta$ on either side of the contact while outside of the angular width $\Delta\theta$ no electrons are emitted. The electron paths are computed by numerically integrating the equation of motion from Eq. 2 and Eq. 3. The transmission *T* between the top and bottom contacts is then computed by counting the fraction of electrons that reach the non-collimating collecting contact, which has contacts 3, 5, and 6 tied together. An image of electron flow is obtained from the simulations, by displaying the transmission change $\Delta T = (T_{tip} - T_{notip})$ *vs*. tip position. The angular width of the experimental electron beam is determined by using the simulations to fit the image intensity data *vs*. *B*.

3. **Results and Discussion**

*3.1. Images of collimated electron flow in zero magnetic field:*

The SGM images of electron flow in Fig. 2a and 2b and simulated images in Fig. 2c and 2d clearly demonstrate that the collimating contact significantly narrows the width of the electron beam emitted into the graphene channel: Figs. 2b and 2d on the right are for a collimating top contact, and Figs. 2a and 2c on the left are for a non-collimating top contact. Experiment and theory agree quite well. In each image, the red regions are where the tip has scattered electrons away from their original trajectories, providing a direct image of the electron flow. The orange lines on the top and bottom of each image show the shape of the emitting and collecting contacts. Comparing images Figs. 2b and 2d for the collimating contact with the non-collimating case in Figs. 2a and 2c, it can be seen that the electron beam from collimating contact is significantly narrower. All these images were obtained at electron density *n* = 1.08 x 10$^{12}$ cm$^{-2}$ and *B* = 0 T.

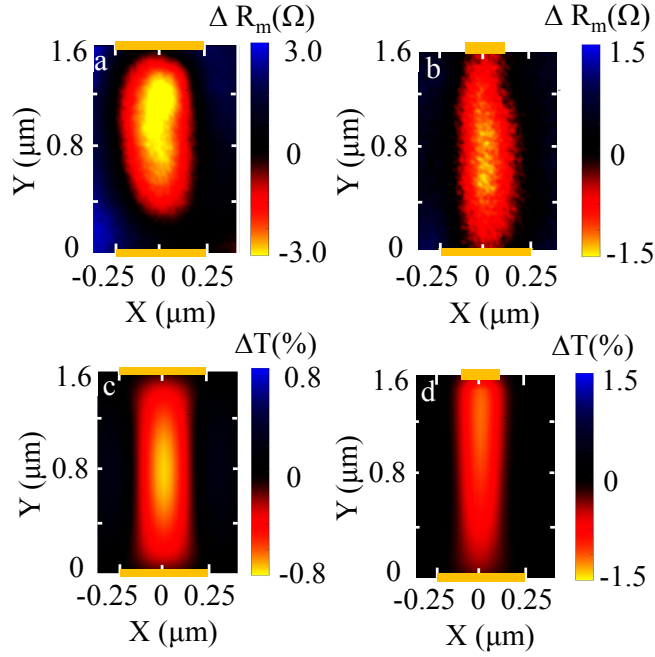

Figure 2: (a) SGM image of electron flow from the non-collimating top contact to the non-collimating bottom contact (Fig. 1a). (b) SGM image of the electron flow when the collimation is turned on in the top contact (see text) (Fig. 1(b)). (c) Simulated image of electron flow from the non-collimating top contact to the non-collimating bottom contact. (d) Simulated image of electron flow when collimation is turned on in the top contact. All images at zero magnetic field $B = 0$ and electron density $n = 1.08 \times 10^{12}$ cm$^{-2}$. The orange bars on the top and bottom of each image show the contact

## 3.2 Angular distribution of the electron beam:

To measure the angular distribution of the electron beam from the collimating contact, a perpendicular magnetic field was applied to bend electron paths away from their original direction so they tend to miss the collecting contact. Figures 3a and 3b show SGM images of electron flow for a non-collimating and a collimating top contact, respectively, tiled against the magnetic field $B$ and electron density $n$. The electron flow shows a strong dependence on $B$. For the non-collimating top contact (Fig. 3a), the electron flow persists up to much higher fields $B = 0.19$ T than for the collimating top contact $B = 0.05$ T. These data show the angular width of the electron beam emitted by the collimating contact is much sharper than the angular width for non-collimating contact. A quantitative analysis is given below.

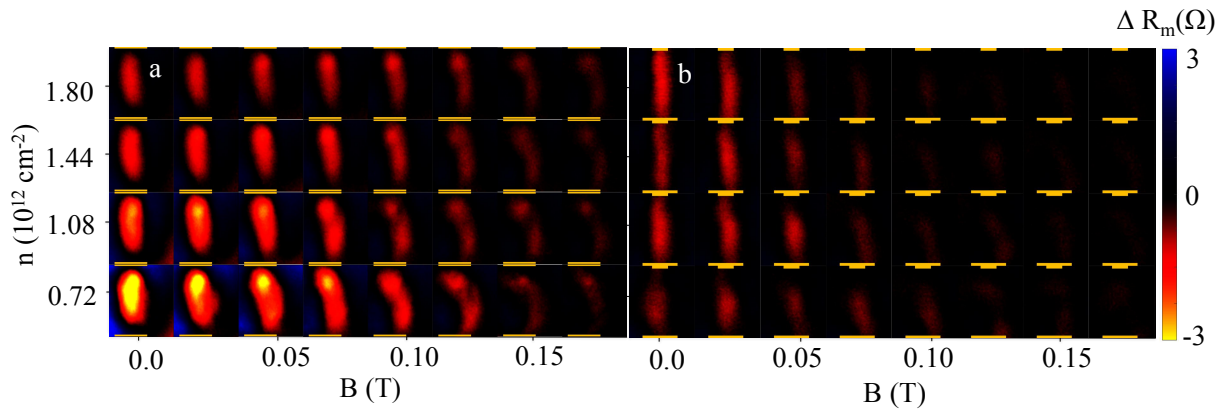

Figure 3: A magnetic field was used to measure the angular width of electron beam emitted into the contact - the field bends their trajectories into cyclotron orbits and causes them to miss the collecting contact. Tiled plot of experimental SGM images of (a) electron flow from the non-collimating top contact to the non-collimating bottom contact and (b) electron flow when collimation is turned on in the top contact. Electron density $n$ is on the vertical axis and magnetic field $B$ on the horizontal axis.

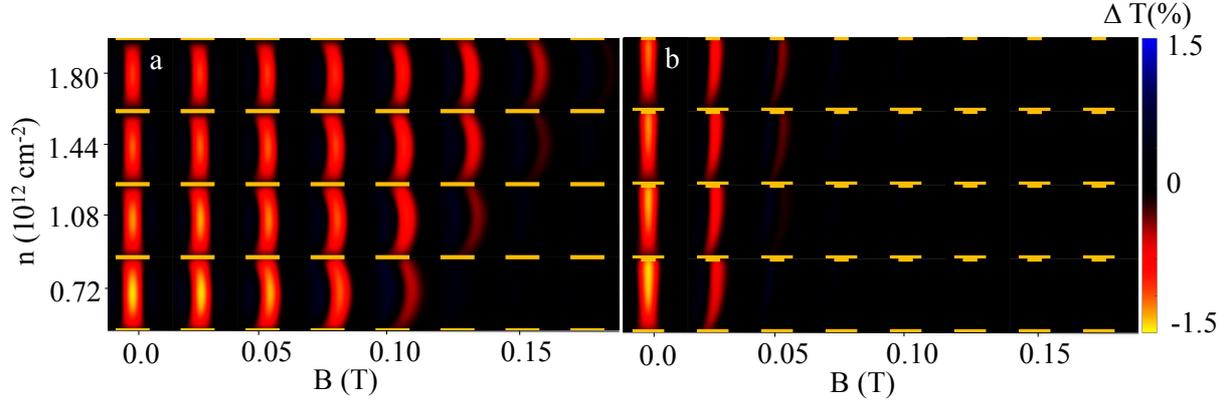

Figure 4: Tiled plot of simulated images of (a) electron flow from the non-collimating top contact to the non-collimating bottom contact and (b) electron flow when collimation is turned on in the top contact. Electron density $n$ is on the vertical axis and magnetic field $B$ on the horizontal axis.

Figure 4 shows ray-tracing simulations of the SGM images in Fig. 3 for both the non-collimating (Fig. 4a) and collimating top contacts (Fig. 4b). The simulations agree well with the experimental images, and confirm that the collimating contacts considerably reduce the angular width $\Delta\theta$ of the emitted electron beam. In these tiled images, the vertical axis represents electron density $n$ while the horizontal-axis represents magnetic field $B$. Similar to the experimental results in Fig. 3, the electron flow between the top and bottom contacts dies off as the magnetic field is increased. Importantly, the electron flow persists to much higher fields for the non-collimating contact (Fig. 4a) than for the collimating contact (Fig. 4b) suggesting that the collimating contact has a much narrower angular width. In addition, the electron flow persists up to higher fields $B$ at higher $n$, confirming that the cyclotron diameter increases with density.

The experimental data (Fig. 3) and simulations (Fig. 4) were used to quantitatively measure the angular width $\Delta\theta$ of the electron beam emitted into the graphene from the collimating top contact (Fig. 5a and 5b) and from the non-collimating contact (Fig. 5c and 5d). The magnitude of electron flow is obtained by taking sum of the transresistance change $\Delta R_m$ at each pixel $(\Sigma \Delta R_m)$ across each experimental SGM image shown in Fig. 3 and sum of the transmission change $\Delta T$ at each pixel $(\Sigma \Delta T)$ for simulated images across Fig. 4. Figures 5a and 5b show the magnitude of electron flow vs. $B$ for the experimental images and simulations for the collimating contact in Figs. 3b and 4b respectively. Figures 5c and 5d show the magnitude of electron flow vs. $B$ for the experimental images and simulations for non-collimating contacts in Figs. 3b and 4b respectively. These comparisons were done at a series of electron densities $n$: the colors purple, yellow, red and blue represent $n = 0.72 \times 10^{12}$ cm$^{-2}$, $n = 1.08 \times 10^{12}$ cm$^{-2}$, $n = 1.44 \times 10^{12}$ cm$^{-2}$ and $n = 1.80 \times 10^{12}$ cm$^{-2}$ respectively. The magnitude of electron flow dies off with $B$ in all cases. For the collimating contact, both the experimental (Fig 5a) and simulated magnitude (Fig 5b) show a rapid decrease in flow as $B$ increases. For the non-collimating contacts, both the experimental (Fig 5c) and simulated magnitude (Fig 5d) show a much slower decrease in flow with $B$.

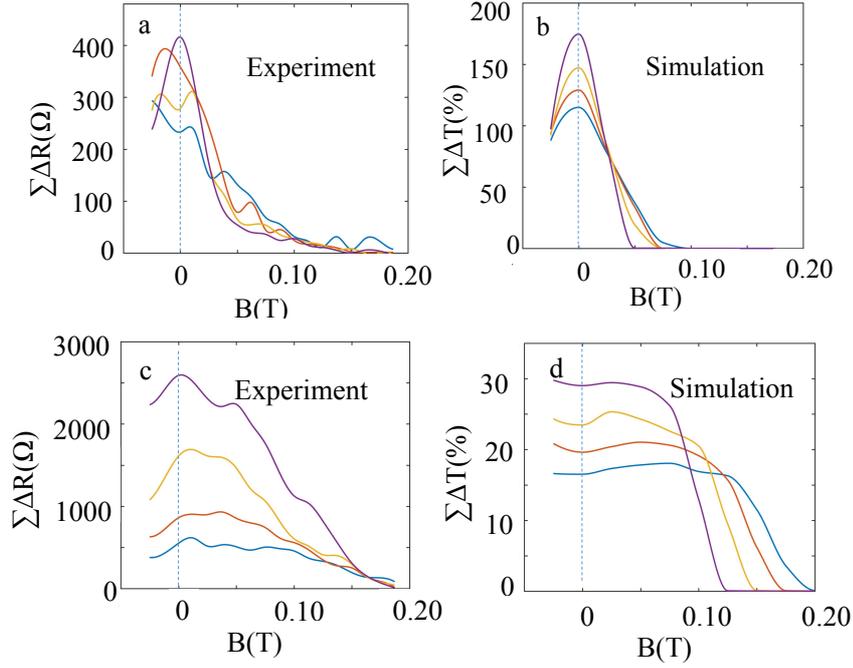

Figure 5: The total magnitude of the electron flow $\sum \Delta R$ between the top and bottom contacts in the SGM images (Fig. 3) vs. $B$ for (a) collimated top contact and (c) non-collimating top contact, together with the total magnitude $\sum \Delta T$ of electron flow in the simulations (Fig.4) vs. $B$ for (b) collimated top contact and (d) non-collimating top contact, shown at the four electron densities: (blue) $n = 0.72 \times 10^{12}$ cm$^{-2}$, (red) $n = 1.08 \times 10^{12}$ cm$^{-2}$, (yellow) $n = 1.44 \times 10^{12}$ cm$^{-2}$, (purple) $n = 1.80 \times 10^{12}$ cm$^{-2}$. The total magnitude $\sum \Delta R$ is the sum of transresistance changes $\Delta R_m$ at all tip locations for experimental images and sum $\sum \Delta T$ of transmission changes $\Delta T$ for simulated images. The electron flow for the collimating contact dies off at much lower $B$ compared non-collimating contact.

The angular width $\Delta \theta$ of the electron beam can be obtained for the collimating and non-collimating top contact by plotting in Fig. 6 the half-width at half-maximum magnetic field HWHM for the experimental (Fig. 5a) and simulated magnitudes (Fig. 5c) vs. the electron density $n$. The HWHM field for the simulations is well described by a linear dependence on density HWHM = a$n$ + b. A fit of the two coefficients to the experimental data for the collimating contact gives a = $2.5 \times 10^{-14}$ Tcm$^2$ and b = $0.1 \times 10^{-2}$ T (red dotted line in Fig. 6). For the non-collimating contact we have a = $5.8 \times 10^{-14}$ Tcm$^2$ and b = $3.9 \times 10^{-2}$ T (blue dotted line in Fig. 6). The fitted parameters a and b determine the angular width $\Delta\theta = 9.2°$ (on either side) of the electron beam exiting the collimating contact and the width $\Delta\theta = 54°$ for the non-collimating contact. The angular width $\Delta\theta$ for the collimating contact is more than five times smaller than for the non-collimating contact. These quantitative results confirm that the design of collimating contact is very effective. Previous transport measurements in graphene that used two constrictions to collimate electron flow show an angular width $\Delta\theta = 9°$ similar to our results [23].

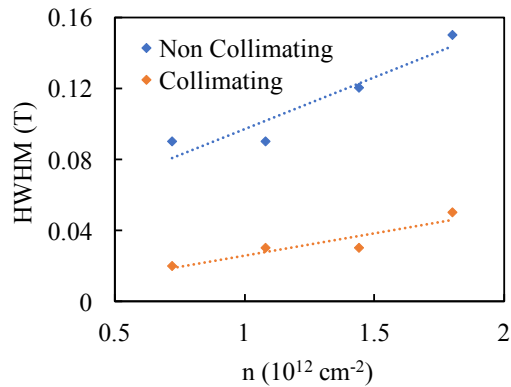

Figure 6: Experimental magnetic field $B$ required to drop the total magnitude of electron flow to half the zero-field value for the collimating top contact (red) (Fig 5a) and for the non-collimating top contact (blue) (Fig 5c). The width for the collimating top contact is approximately four times smaller than for the non-collimating contact, showing that the angular width of the electron beam is narrower.

## 4. Conclusion

By imaging the electron flow with a cooled Scanning Gate Microscope (SGM) we have shown that a collimating contact design based on zigzag side contacts considerably narrows the angular width of the electron beam emitted into the graphene sample. We observe a spatially narrow beam of electron flow with angular width $\Delta\theta = 9.2°$ for the collimating contact which is more than five times narrower than the angular width $\Delta\theta = 54°$ for the non-collimating contact. The ability for a contact to create a narrow electron beam is promising for future experiments on ballistic devices in graphene as well as other atomic layer materials.

## 5. Acknowledgement

The SGM imaging experiments and the ray-tracing simulations were supported by the DOE Office of Basic Energy Sciences, Materials Sciences and Engineering Division, under grant DE-FG02-07ER46422. The graphene sample fabrication was supported by Global Research Laboratory Program (2015K1A1A2033332) through the National Research Foundation of Korea (NRF). Growth of hexagonal boron nitride crystals was supported by the Elemental Strategy Initiative of MEXT, Japan and a Grant-in-Aid for Scientific Research on Innovative Areas No. 2506 "Science of Atomic Layers" from JSPS. Nanofabrication was performed in the Center for Nanoscale Systems (CNS) at Harvard University, a member of the National Nanotechnology Coordinated Infrastructure Network (NNCI), which is supported by the NSF under NSF award ECCS-1541959.